\documentclass[manuscript]{aastex}
\usepackage{graphicx}
\usepackage{rotating}
\usepackage{subfigure}
\usepackage{amsmath}
\usepackage{color}
\usepackage{eucal}
\usepackage{lineno}
\usepackage{natbib}
\linenumbers

\begin{document}

\title{Suppression of Dielectronic Recombination Due to Finite Density Effects}

\author{D. Nikoli\'c\altaffilmark{1}, T.~W.~Gorczyca, K.~T. Korista}
\affil{Western Michigan University, Kalamazoo, MI, USA}
%\email{dragan.nikolic@wmich.edu}

\author{G.~J. Ferland}
\affil{University of Kentucky, Lexington, KY, USA}
%\email{kirk.korista@wmich.edu}

\and

\author{N.~R. Badnell}
\affil{University of Strathclyde, Glasgow, UK}
%\email{badnell@phys.strath.ac.uk}

\altaffiltext{1}{Department of Mechanical Engineering, University of Alberta, Edmonton AB, Canada}

\begin{abstract}
We have developed a general model for determining density-dependent effective dielectronic recombination (DR)
rate coefficients in order to explore finite-density effects on the ionization balance of plasmas.
Our model consists of multiplying by a suppression factor those highly-accurate total zero-density DR rate coefficients
which have been produced from state-of-the-art theoretical calculations and which have been benchmarked by experiment.
The suppression factor is based-upon earlier detailed collision-radiative calculations
which were made for a wide range of ions at various densities and  temperatures, but used a simplified treatment of DR.
A general suppression formula is then developed as a function of isoelectronic sequence, charge, density, and temperature.
These density-dependent effective DR rate coefficients are then used in the plasma simulation code Cloudy to
compute ionization balance curves for both collisionally ionized and photoionized plasmas at very low ($n_{\rm{e}}=1~\rm{cm}^{-3}$)
and finite ($n_{\rm{e}}=10^{10}~\rm{cm}^{-3}$) densities.  We find that the denser case is significantly more ionized due to
suppression of DR, warranting further studies of density effects on DR by detailed collisional-radiative calculations which
utilize state-of-the-art partial DR rate coefficients. This is expected to impact the predictions of the ionization balance
in denser cosmic gases such as those found in nova and supernova shells, accretion disks, and the broad emission line regions
in active galactic nuclei.
\end{abstract}


%===============================%
%===% SECTION: INTRODUCTION %===%
%===============================
\section{Introduction}
\label{intro}

Astronomical emission or absorption sources have an enormous range of densities.  Two examples include the
intergalactic medium, with $n_{\rm{e}} \sim 10^{-4}\ {\rm cm}^{-3}$, and the broad emission-line regions of Active Galactic Nuclei,
with $n_{\rm{e}} \sim 10^{10}\ {\rm cm}^{-3}$.  The gas producing the spectrum is not in thermodynamic equilibrium
\citep{osterbrockferland},  so microphysical processes determine the physical conditions.

The two common cases encountered for ionization are photoionization and collisional (e.g., electron-impact) ionization.
In both cases, ions are recombined by dielectronic and radiative recombination, with dielectronic recombination (DR) usually
the dominant process for elements heavier than helium.  Databases give ionization and recombination rates that are the sum
of several contributing processes.  Examples include \citet{voronov}  for electron impact ionization, \cite{verner} for
photoionization, and the DR project \citep{drproject} for dielectronic recombination and \cite{rrproject} for radiative recombination;
 it is these latter data \footnote{\tt http://amdpp.phys.strath.ac.uk/tamoc/DATA/}
which will be of primary interest to us in the present study.

The collisional ionization and recombination rate coefficients used in astrophysics are frequently assumed to depend on
temperature but to have no density dependence.  The rigorous treatment of density dependent ionization and recombination rate coefficients
is via collisional-radiative modeling. This was introduced by \citet{bates} for radiative recombination only and
extended to treat the much more complex case of dielectronic recombination by \citet{burgsum}.
Summers applied their techniques to determine density dependent ionization and recombination rate coefficients, and the
consequential ionization balance for collisional plasmas, for H-like thru Ar-like ions.
Graphical results were presented for the elements C, O and Ne \citep{summers1} and then N, Mg and Si \citep{summers2}.
Reduced temperatures and densities were used so as to enable easy interpolation for other elements.
Tables of such recombination rate coefficients were made available only via a Laboratory Report --- \citet{summersRAL} ---
due to their voluminous nature at that point in history. The `difficulty' in utilizing this pioneering data
led to some modelers attempting to develop simplified approaches. For example, \citet{jordan} used an approach which was based on
truncating the zero-density DR sum over Rydberg states using a simple density dependent cut-off which itself was based on
early collisional-radiative calculations by \citet{burgsum};
a suppression factor was formed from its ratio to the zero-density value and then used more generally.
Also, \citet{davidson} simplified the collisional-radiative approach of \citet{burgsum} and, using hydrogenic atomic data,
determined suppression factors for Li-like \ion{C}{4} and \ion{O}{6}.
New calculations for \ion{C}{4} were made by \citet{badnell} utilizing more advanced (generalized) collisional-radiative modeling \citep{sumhoop}
and much improved atomic data at collisional plasma temperatures (see the references in \citet{badnell}).

All of the above works were for electron collisional plasmas and used rather basic DR data (excluding \citet{badnell})
as epitomized in the \citet{burgess}
General Formula, viz. a common dipole transition for dielectronic capture, autoionization, and radiative stabilization.
The purpose of the present paper is to explore density suppression of DR in photoionized plasmas, and within collisional
plasmas, using state-of-the art DR data which takes account of a myriad of pathways not feasible in the early works above,
but which has been shown to be necessary by comparison with experiment. We wish to gain a broad overview utilizing the large
test-suite maintained by the plasma simulation code Cloudy. We utilize an approach to DR suppression which is motivated
initially by the detailed  collisional-radiative results given in \citet{badnell} for \ion{C}{4} at $T=10^5$ K, along with known scalings
to all temperatures, charges,  and densities.  Using these results as a guideline, a more general
suppression formula is then determined by fitting to suppression results from extensive detailed collisional-radiative calculations
\citep{summersRAL} for a wide range of ions at several densities and (high) temperatures.
Additional modifications are then introduced to account for low temperature DR.

%Whereas a generalized collisional-radiative modeling approach would provide the most definitive suppression results, such data
%is mainly available for the light elements of interest to magnetic fusion and is based on $LS$-coupling DR data. Such data
%tends to be a drastic underestimate at photoionized plasma temperatures, although they do much better at the higher temperatures
%applicable to collisional plasmas. Collisional-radiative modeling with the intermediate coupling DR data of \cite{badnell}
%is in its early stages. Nevertheless, given the increasing accuracy both of modeling and observation,
%it is desirable to ascertain, within the context of astrophysical importance, whether or not those more elaborate calculations
%are necessary for plasma simulations at densities of up to $n_{\rm{e}} \sim 10^{10}\ {\rm cm}^{-3}$.

The outline of the rest of the paper is as follows: in the next section we describe the DR suppression model we use;
we then apply this suppression to the zero-density DR data, and use the resultant density-dependent DR data in
Cloudy to determine the ionization distribution produced under photoionized and collisional ionization equilibrium at
low and moderate densities.

\section{Generalized Density Suppression Model}
\label{Sec:Suppress}

We use the following approach, detailed more fully in the subsections below.
First, the high-temperature collisional-radiative modeling results of \citet{badnell} for DR suppression
in \ion{C}{4} are parameterized by a pseudo-Voigt profile to study the qualitative behavior of suppression as a function of density and temperature.  Next, this formulation is then
used as a guideline for developing a more comprehensive suppression formula which is obtained by
fitting to collisional radiative data for various isoelectronic sequences, ionic charges, densities, and temperatures \citep{summersRAL}.
Lastly, the suppression formulation is extended to low-temperatures according to the nature of the
sequence-specific DR.

\subsection{High-Temperature Suppression for Li-like \ion{C}{4}}

We begin by considering DR of Li-like \ion{C}{4}, for which the density dependent total DR rate coefficient,
and therefore the suppression factor, has been computed rigorously within a collisional-radiative modeling approach \citep{badnell}.

In the electron collisional ionization case, because of the consequential high temperature of peak abundance,
dielectronic recombination occurs mainly through energetically high-lying autoionizing
states (via dipole core-excitations) for which radiative stabilization is by the core electron
into final states just below the ionization limit:
\begin{eqnarray}
e^- + 1s^22s  \rightarrow &
             1s^22pnl  \rightarrow &  1s^22snl + h\nu\ .
	          \label{eq1}
\end{eqnarray}
 In the zero-density limit, the intermediate $1s^22snl$ states can only decay further via radiative cascading until the
$1s^22s^2$  final recombined ground state is reached, thereby completing the DR process:
\begin{eqnarray}
1s^22snl \longrightarrow & 1s^22sn^\prime l^\prime+ h\nu_1  \rightarrow ... \rightarrow & 1s^22s^2 + h\nu_1+h\nu_2+... \label{eqradstab}
\end{eqnarray}
For finite electron densities $n_{\rm{e}}$, on the other hand,
there is also the possibility for reionization via electron impact, either directly or stepwise,
\begin{eqnarray}
e^-  + 1s^22snl  \longrightarrow  & 1s^2 2sn^\prime l^\prime+ e^-  \rightarrow ... \rightarrow & 1s^22s + e^- + e^- \ ,\label{eqreion}
\end{eqnarray}
and the probability of the latter pathway is proportional to the electron density $n_{\rm{e}}$.
Because of this alternative reionization pathway at finite densities, the {\em effective} DR rate coefficient
$\alpha^{\rm{eff}}_{\rm{DR}}(n_{\rm{e}},T)$
is thus  {\em suppressed} from the zero-density value $\alpha_{\rm{DR}}(T)$ by a density-dependent suppression factor $S(n_{\rm{e}},T)$:
\begin{eqnarray}
\alpha^{\rm{eff}}_{\rm{DR}}(n_{\rm{e}},T) & \equiv & S(n_{\rm{e}},T)\alpha_{\rm{DR}}(T)\ .
\label{eqsuppress}
\end{eqnarray}

\begin{figure}[!hbtp]
\centering
\includegraphics[width=\textwidth]{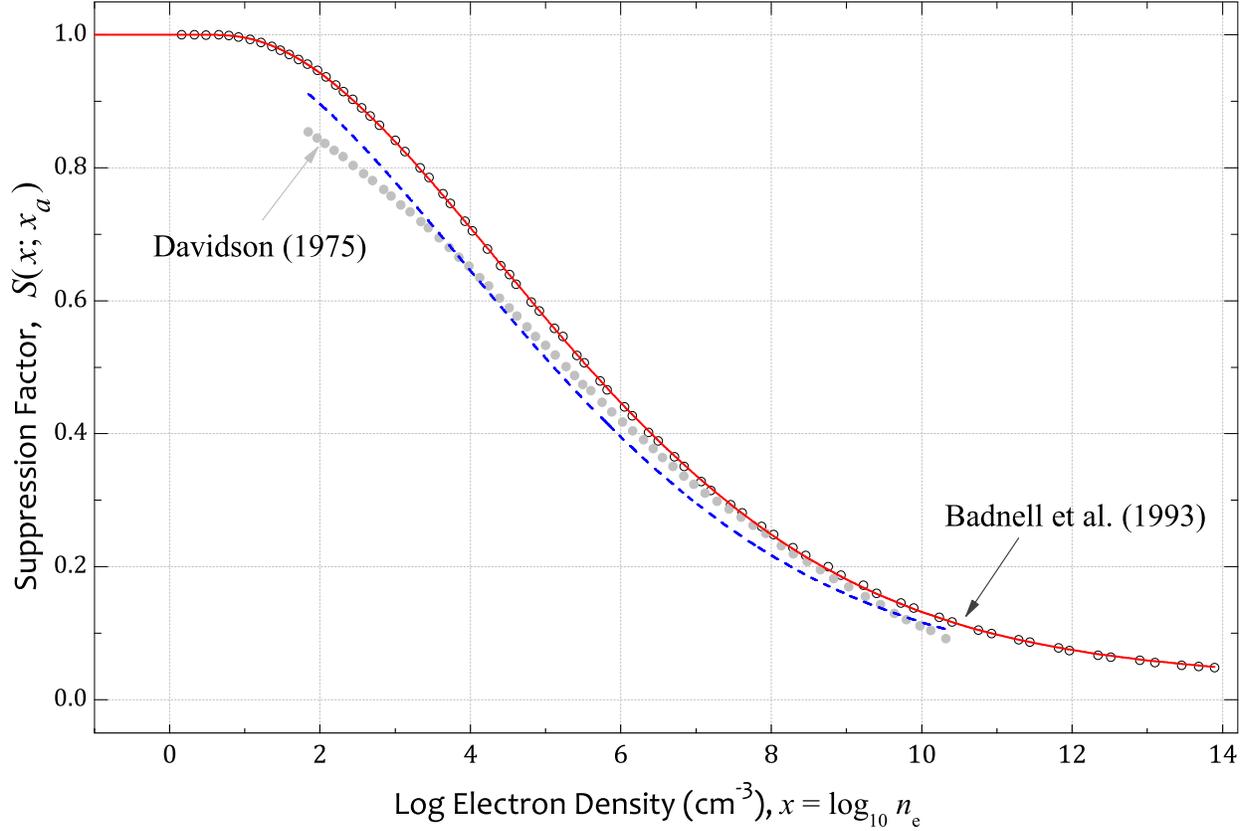}
\caption{\label{fig1}
Pseudo-Voigt fit of the suppression factor for \ion{C}{4}, as given in Eq.~\ref{eqsuppression} with a scaled activation density as given by Eq.~\ref{eqxa},
shown for two different temperatures. The red solid curve shows that the parameterization
 for $T=1\times10^{5}$~K, corresponding to an activation density of $x_{\rm{a}}=0.608$
 (with $\mu=0.372$ and $w=4.969$),
is in close agreement with the actual data of \cite{badnell} (open circles).
The  blue dashed curve is the parameterization
 for $T=1.5\times10^{4}$~K, using instead an activation density of $x_{\rm{a}}=0.196$ (and the same $\mu$ and $w$),
and giving satisfactory agreement with the data of \cite{davidson} (solid circles).
}
\end{figure}

From the earlier detailed studies of  \citet{davidson} and \citet{badnell}, the suppression factor is found to remain unity, corresponding to
zero suppression, at lower densities until a certain {\em activation} density $n_{\rm{e},\rm{a}}$ is reached, beyond which this factor decreases
exponentially from unity with increasing density. We have found that this suppression factor, as a function of the dimensionless log
density parameter $x=\log_{10}n_{\rm{e}}$,  can be modeled quite effectively by a pseudo-Voigt profile~\citep{Wertheim:1974} --- a weighted
mixture $\mu$ of Lorentzian and Gaussian profiles of widths $w$ for densities above the activation density $x_{\rm{a}}=\log_{10}n_{\rm{e},\rm{a}}$:
\begin{equation}
   S(x;x_{\rm{a}}) = \left\{
    \begin{matrix}
    1  &  x \le x_{\rm{a}} \\
    \mu \left[\frac{1}{1+(\frac{x-x_{\rm{a}}}{w})^2}\right]
    + (1 - \mu)\left[{\rm e}^{-(\frac{x-x_{\rm{a}}}{w/\sqrt{\ln 2}})^{2}}\right]  &  x \geq x_{\rm{a}}
    \end{matrix}\right. \ .
    \label{eqsuppression}
\end{equation}
Fitting this expression to the suppression factor of \citet{badnell}
for \ion{C}{4} (which was computed at $T=10^5$ K) yielded the values $\mu=0.372$, $w=4.969$, and $x_{\rm{a}}=0.608$,
and this parameterization formula is found to be accurate to within 5\% for all densities considered (see Fig.~\ref{fig1}).

\subsection{Generalized High-Temperature Suppression Formula}
\label{BurgessHT}
Given the suppression formula for Li-like \ion{C}{4}, corresponding to ionic charge $q_0=3$ and temperature $T_0=10^5$ K, we wish to
generalize  this expression to other Li-like ions of charge $q$ and (high) $T$ according to the following qualitative guidelines.
It is well known that density effects scale as $q^7$ --- see \citet{bates} and \citet{burgsum}.
The activation density is attained when the reionization rate in Eq.~\ref{eqreion}, which depends linearly on the density, becomes comparable
to the radiative stabilization rate in Eq.~\ref{eqradstab}.  The radiative rate is independent of density and temperature, but scales
with charge as $A_r\sim q^4$, whereas the electron-impact ionization rate
depends on all three, viz. $n_{\rm{e}}\alpha_{\rm{eII}}\sim n_{\rm{e}} q^{-3}T^{-1/2}$.
An initial suggestion is that the activation density is attained when these two are approximately equal, i.e.,
\begin{eqnarray}
n_{\rm{e},\rm{a}}q^{-3}T^{-1/2}\sim q^4\ ,
\end{eqnarray}
indicating that the activation density should scale as $ n_{\rm{e},\rm{a}}\sim q^7T^{1/2}$,
if the above qualitative discussion holds.
The log activation density for all $q$ and $T$ might therefore be
expected to obey the scaling relationship
\begin{eqnarray}
x_{\rm{a}}(q,T) & = & x_{\rm{a}}(q_0,T_0) + log_{10}\left[\left(\frac{q}{q_0}\right)^7\left(\frac{T}{T_0}\right)^{1/2}\right]\ , \label{eqxa}
\end{eqnarray}
where $x_{\rm{a}}(q_0,T_0)=0.608$, $q_0=3$, and $T_0=10^5$ K are the (log) activation density, the charge, and the temperature for the
\ion{C}{4} case treated by \cite{badnell}.
We note that this expression, when applied to Li-like \ion{O}{6}, gives
an increase in the activation density by a factor of $(5/3)^7 = 35.7$, in agreement with the approximate factor of 40 found by \citet{davidson}.
Furthermore, when scaled in temperature, the formula gives fairly good agreement with the
suppression results of \citet{davidson} for \ion{C}{4} at $T=1.5\times 10^4$ K
(see Fig.~\ref{fig1}).

\subsubsection{Fit to the Collisional Radiative Data}
\label{cr}

The preceding treatment reasonably extends the \ion{C}{4} suppression factor at $10^5$~K to other high temperatures and to other Li-like ions.
However, we need suppression factors applicable to all ionization stages of all elements up to at least Zn for a general implementation
within Cloudy. Unfortunately, detailed collisional-radiative modeling data with state-of-the-art DR data is still rather limited.
However, extensive tables of effective recombination rate coefficients have been computed by  \cite{summersRAL}
for a wide variety of isoelectronic sequences, charge-states, temperatures, and densities. The treatment of DR there is somewhat
simplified, but we only require the {\it ratio} of finite- to zero-density rate coefficients to determine the suppression factor.
We then combine this ratio with our state-of-the-art zero density DR rate coefficients again for use within Cloudy.
This ratio is much less sensitive to the specific treatment of DR.

The rather simplistic scaling formula in Eq.~\ref{eqxa} was found to be inadequate when applied to the extensive
tabulation of suppression factors found in \cite{summersRAL}.  Instead, in order
to fit the data accurately, a more generalized formula was arrived at,
where a  pseudo-Gaussian, corresponding to $\mu=0$ in Eq.~\ref{eqsuppression}, was more appropriate,
\begin{equation}
   S^N(x;q,T) = \left\{
    \begin{matrix}
    1  &  x \le  x_a(q,T,N)\\
    {\rm e}^{-(\frac{x-x_a(q,T,N)}{w/\sqrt{\ln 2}})^{2}}  &  x \geq x_a(q,T,N)
    \end{matrix}\right. \ .
    \label{eqsuppression2}
\end{equation}
Furthermore,
the activation density was found to be best represented by the function
\begin{eqnarray}
x_a(q,T,N) & = & x_{a}^{0} + log_{10}\left[\left(\frac{q}{q_0(q,N)}\right)^{7}\left(\frac{T}{T_0(q,N)}\right)^{1/2}\right]\ , \label{eqxanew}
\end{eqnarray}
where the variables $q_0(q,N)$ and $T_0(q,N)$ are taken to be functions of the charge $q$ and the isoelectronic sequence, labeled by $N$.
A fit of the suppression factors of \citet{summersRAL} for all ions yielded a global (log) activation density $x_{a}^{0}=10.1821$ and more complicated
expressions for the zero-point temperature $T_0$ and charge $q_0$.
These were found to depend on both the  ionic charge $q$ and the isoelectronic sequence $N$ viz.
\begin{eqnarray}
T_0(q,N) &= &5\times10^{4}\,[q_0(q,N)]^{2} \label{t0}
\end{eqnarray}
and
\begin{eqnarray}
q_0(q,N) & = &  (1 - \sqrt{2/3q})A(N)/\sqrt{q}\ , \label{q0}
\end{eqnarray}
where
\begin{eqnarray}
A(N) & = &  12 + 10N_{1} + \frac{10N_{1}-2N_{2}}{N_{1}-N_{2}}(N-N_{1}) \label{AN}
\end{eqnarray}
depends on the isoelectronic sequence in the periodic table according to the specification of the parameters
\begin{eqnarray}
(N_{1},N_{2}) & = &
\begin{pmatrix}
  (3,10)  & N\in 2^{nd} \, \rm{row}  &   & (37,54)  & N\in 5^{th} \, \rm{row} \\
  (11,18) & N\in 3^{rd} \, \rm{row}  &   & (55,86)  & N\in 6^{th} \, \rm{row} \\
  (19,36) & N\in 4^{th} \, \rm{row}  &   & (87,118) & N\in 7^{th} \, \rm{row} \\
\end{pmatrix}\ .
\label{N1N2}
\end{eqnarray}
However, even this rather complicated parameterization was inadequate for the lower isoelectronic sequences $N\le 5$, and for these we explicitly list the
optimal values for $A(N)$ in Table~\ref{TableXX}.
Furthermore, at electron temperatures and/or ionic charges for which the $q$-scaled temperature $\theta \equiv T/q^2$  was very low ($\theta\leq 2.5\times10^{4}$ K),
 a further modification to the coefficients $A(N)$ for $N\le 5$ is necessary
 in that the values in Table~\ref{TableXX} should be multiplied by a factor of two.

\begin{deluxetable}{rcc}
%\tabletypesize{\scriptsize}
\tablecaption{Modified $A(N)$ coefficients from Eq.~(\ref{AN}).
\label{TableXX}}
\tablewidth{0pt}
\tablehead{
\colhead{Sequence} & \colhead{ $N$} & \colhead{$A(N)^{\dag}$}
}
\startdata
 H-like    & 1 & 16 \\
He-like    & 2 & 18 \\
Li-like    & 3 & 66 \\
Be-like    & 4 & 66 \\
 B-like    & 5 & 52 \\
\hline
\multicolumn{3}{l}{${}^{\dag}$ {\footnotesize These must be multiplied by 2.0 if $\theta \leq 2.5\times10^{4}$ K} } \\
\hline\enddata
\end{deluxetable}

\begin{figure}[!hbtp]
\begin{tabular}{ll}
\includegraphics[width=3.2in]{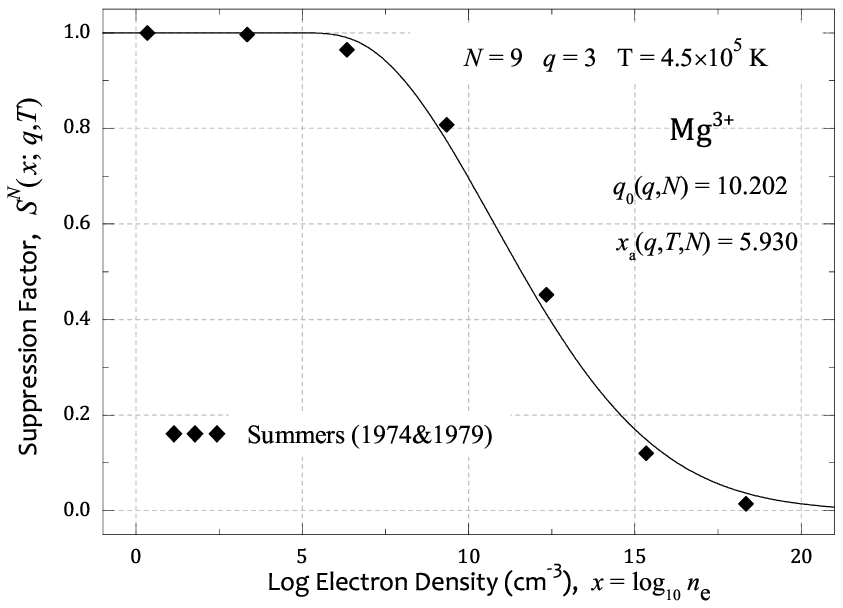}
&
\includegraphics[width=3.2in]{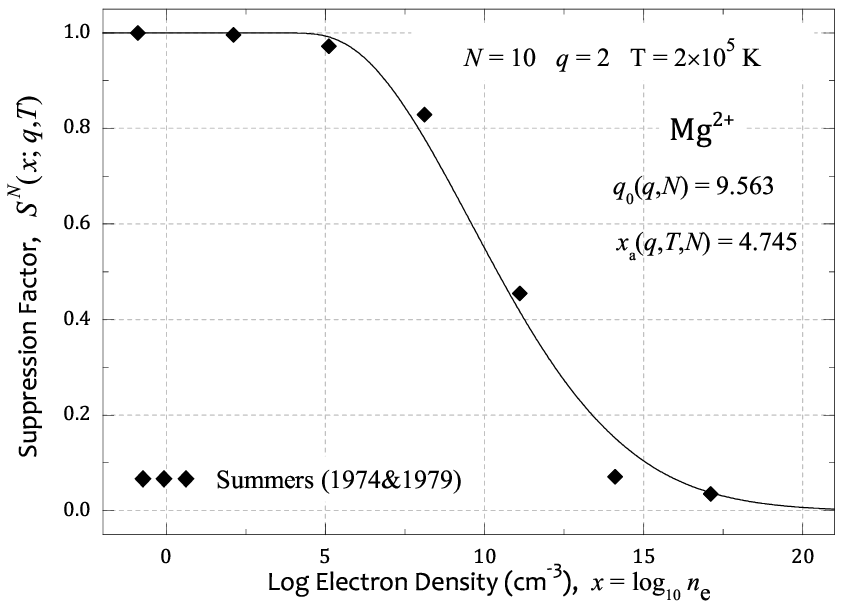}
\\
\includegraphics[width=3.2in]{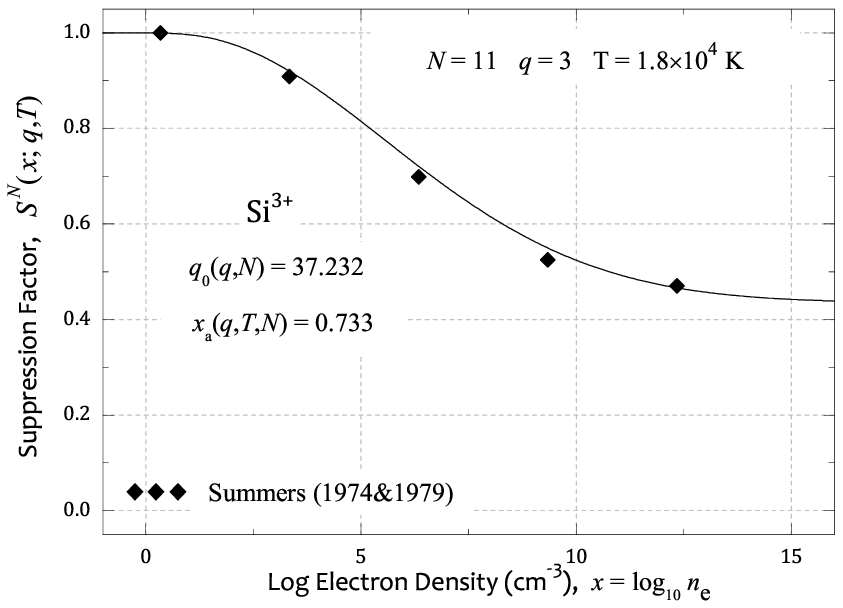}
&
\includegraphics[width=3.2in]{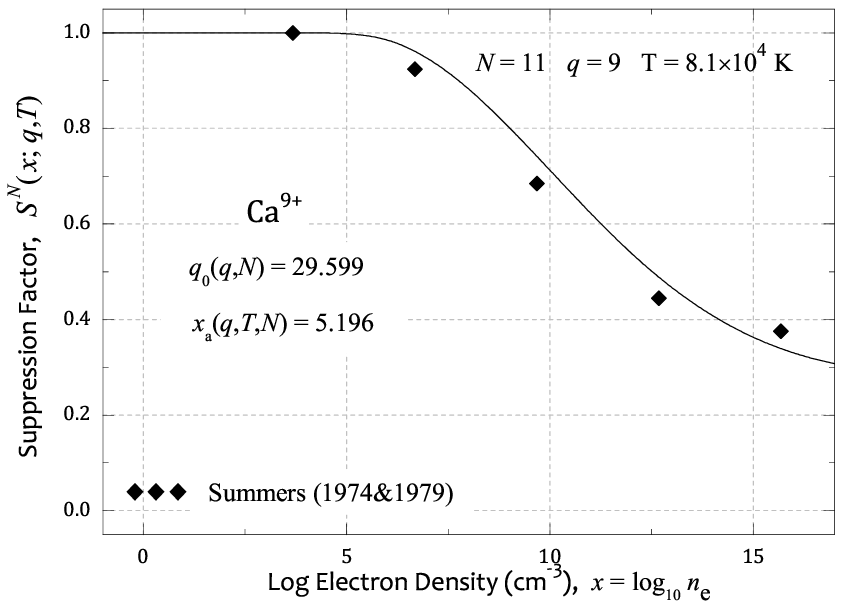}
\\
\includegraphics[width=3.2in]{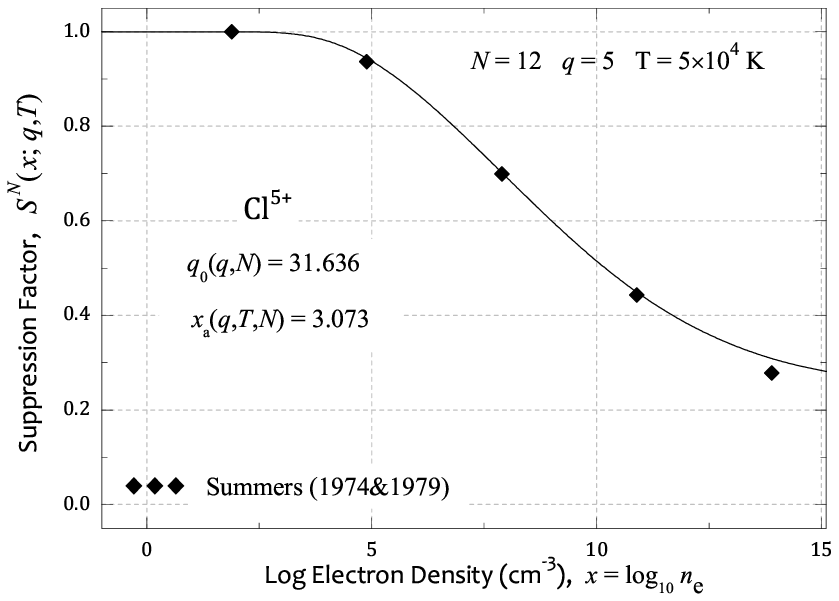}
&
\includegraphics[width=3.2in]{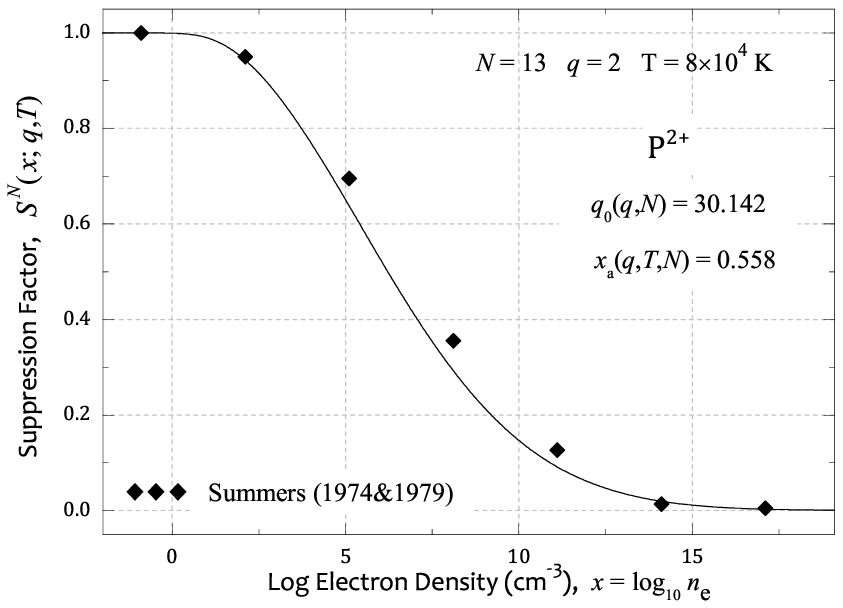}
\end{tabular}
\caption{ A comparison between the present parameterized suppression factor and the collisional radiative results of \citet{summersRAL} for a sample of ions and temperatures, as a function of density.
\label{figsurvey}
}
\end{figure}

\begin{figure}[!hbtp]
\centering
\includegraphics[width=0.7\textwidth]{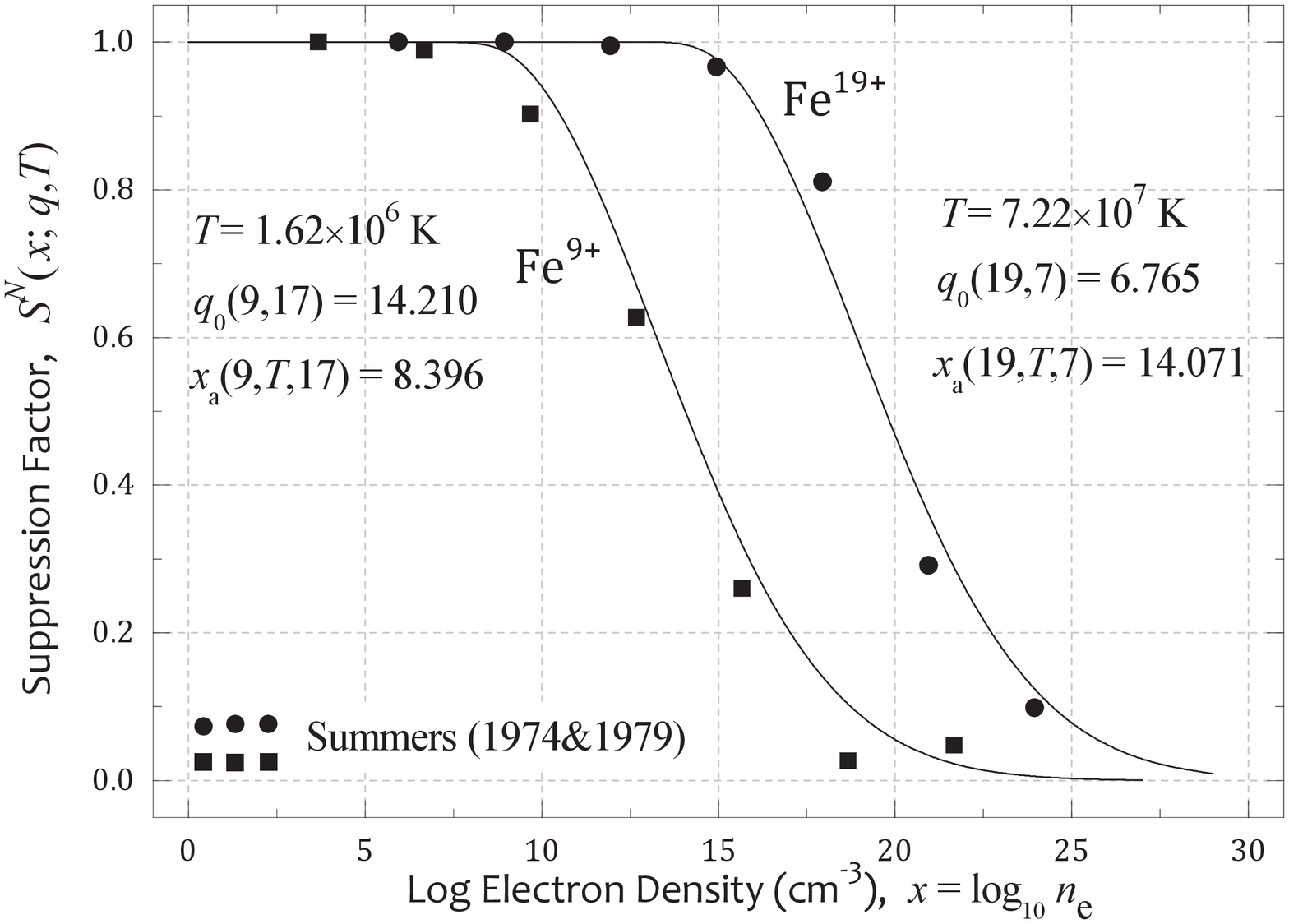}
\includegraphics[width=0.7\textwidth]{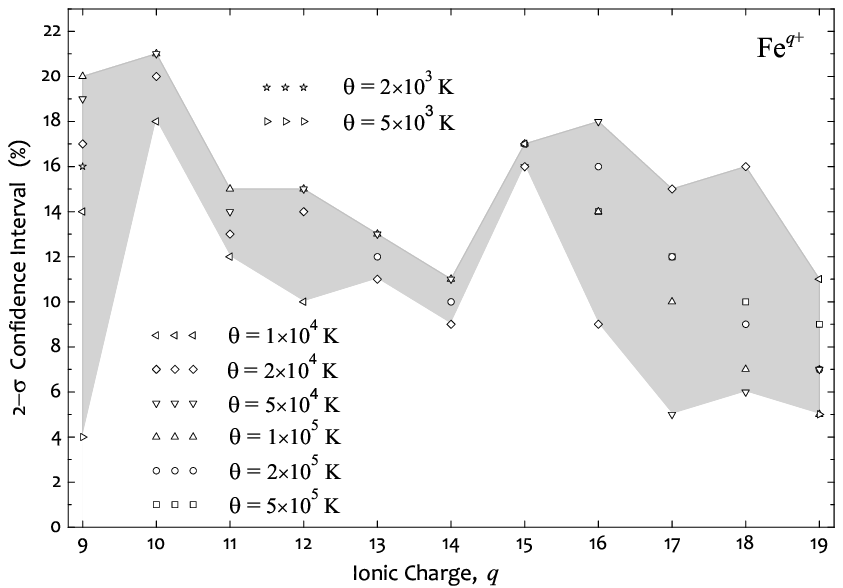}
\caption{\label{figiron}
Agreement between the suppression curve of Eq.~\ref{eqxanew} and the  \citet{summersRAL} data for all iron ions Fe$^{q+}$, $q=9-19$. The upper panel shows the detailed level of agreement of the two end cases, Fe$^{9+}$ and Fe$^{19+}$.
The lower panel shows  the $2-\sigma$ (95.4\%) confidence level as a function of charge state; this means that  95.4\% of all density data points in the \citet{summersRAL} data, for the given charge and temperature, are within that percentage of the prediction from Eq.~\ref{eqsuppression2}.
The symbols denote different values of the scaled temperature $\theta=T/q^2$.
}
\end{figure}

The above final formulation, which consists of the use of Eq.~\ref{eqsuppression2}, with $\mu=0$,
$w=5.64548$, and a rather complicated activation density given by Eqs.~\ref{eqxanew}, \ref{t0}, \ref{q0}, \ref{AN}, and \ref{N1N2}, with $x_{a}^{0}=10.1821$, has been found to model the entire database of ions, temperatures, and densities considered in the \citet{summersRAL} data fairly well.
To illustrate the general level of agreement over a large range of ions and environments,
we compare our parameterized model formulation to the actual suppression data from that report \citep{summersRAL} for a few selected cases in Fig.~\ref{figsurvey}.
In order to quantify more fully the extent of agreement, we focus on the case of iron ions,
for which we study  density effects on ionization balance determination in the next section.  A comparison is shown in Fig.~\ref{figiron} between our predicted suppression factors and the data from the \citet{summersRAL} report.
It is seen that our model fits that data to within 21\% for all densities, temperatures, and ionic stages reported \citep{summersRAL}.
More broadly, we have applied a similar $2-\sigma$ analysis to {\em all} ions in that report, at all temperatures and densities, and find the same agreement (20-26\% confidence level).

\begin{figure}[!hbtp]
\centering
\includegraphics[width=\textwidth]{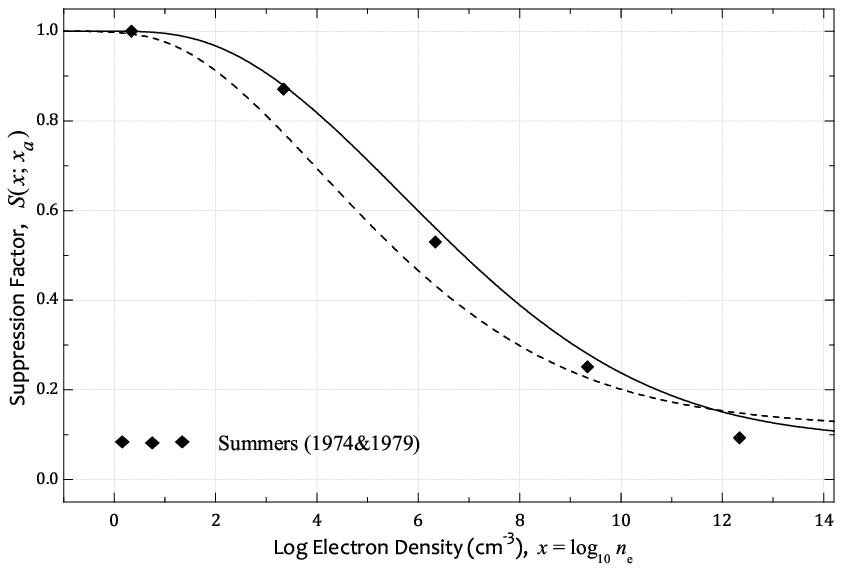}
\caption{\label{figcivnew}
A comparison between the final suppression factor of Eq.~\ref{eqsuppression2} (solid line), corresponding to a pseudo-Gaussian profile with activation density $x_a=0.8314$ ($q_0=40.284$) and width $w=5.64548$,
the \citet{summersRAL} data points (solid diamonds), and the original formulation of Eq.~\ref{eqsuppression} (dashed line),
corresponding to a pseudo-Voigt profile with activation density $x_a=0.608$, width $w=4.696$, and mixture coefficient $\mu=0.372$.
The temperature $T=1\times10^{5}$~K is the same as in Fig.~\ref{fig1}.
}
\end{figure}

Lastly, it is of interest to investigate how our final suppression factor in Eq.~\ref{eqsuppression2} compares to our
original, motivating, formulation of Eq.~\ref{eqsuppression2} for \ion{C}{4}, shown in Fig.~\ref{figcivnew}.
There is generally good qualitative agreement. However, it is seen that the original formulation, based on the \citet{badnell} results,
shows a somewhat stronger suppression effect up to $x\approx 11$. This is likely due to the more accurate treatment of
the partial DR data of \citet{badnell} entering the collisional-radiative modeling, although some difference due to the collisional-radiative
modeling itself may also be present. This indicates that even collisional plasmas require collisional-radiative modeling with state-of-the-art DR data.
The stronger suppression tails-off at $x\gtrsim 11$ as three-body recombination starts to become relevant and which, at even higher densities
(not shown), causes the suppression factor to rise (since it is a ratio of effective recombination rate coefficients, i.e. includes three-body
recombination.)

\subsection{Suppression Formula at Low Temperatures.}
\label{NSFSLT}

The preceding formulation was based on the suppression factor found by \citet{summersRAL} for electron collisionally ionized plasmas,
i.e.,  at higher temperatures, where DR is dominated by high-$n$ resonances attached to a dipole-allowed core excited state.
In photoionization equilibrium, however, the temperature at which a given ion forms is substantially
smaller than that found in the electron collisional case.
 Due to the lower kinetic temperatures, DR occurs mainly through energetically low-lying autoionizing states, often
via non-dipole core-excitations for which radiative stabilization is by the (outer) Rydberg electron. These
states are not, in general, as susceptible to density suppression as their high-$n$ counterparts,  and so
it may be necessary to modify the preceding suppression formulation.

We first consider sequences with partially-occupied $p$-subshells in the ground state, which includes the
B-like $2p(^2P_{1/2,3/2})$,
C-like $2p^2(^3P_{0,1,2})$,
O-like $2p^4(^3P_{0,1,2})$,
F-like $2p^5(^2P_{3/2,1/2})$,
Al-like $3p(^2P_{1/2,3/2})$,
Si-like $3p^2(^3P_{0,1,2})$,
S-like $3p^4(^3P_{0,1,2})$,
and Cl-like $3p^5(^2P_{3/2,1/2})$ systems.
For these sequences, there is fine-structure splitting in the ground state and a correspondingly small excitation energy,
$\epsilon_N$, giving dielectronic capture into high principal quantum numbers
(because of the Rydberg relation $q^2/n^2\le \epsilon_N$).
Stabilization is via $n\rightarrow n'$ transitions and the recombined final state is built upon an excited parent.
Ultimately, it is the strength of collisional coupling of this final state with the continuum which
determines whether recombination or ionization prevails. As the density increases, collisional LTE
extends further down the energy spectrum.
It is difficult to give a general statement about the position of such final states relative to the ionization limit.
So, we assume a worst case scenario, i.e., that such states are subject to suppression, and we
use the preceding suppression formula. If density effects are found to be small in photoionized plasmas then
this is likely sufficient. If they appear to be significant then a more detailed treatment based on collisional-radiative
modeling will be needed.
Thus, for these systems,  we retain the same  suppression
formula developed above, that is, $S^N(x,q,T)=S(x,x_a(q,T))$ for $N=\left\{5,6,8,9,13,14,16,17\right\}$, and for all $q$ and $T$.

For the hydrogenic and the closed-shell He-like and Ne-like cases, on the other hand,
the excitations proceed via an increase in core principal quantum number ---
$1s\rightarrow 2s$ or $\left\{2s,2p\right\}\rightarrow\left\{3s,3p,3d\right\}$ ---
giving the dominant dielectronic capture into the low-$n<10$ resonances.  Even following core radiative
stabilization, these low-lying states are impervious to collisional reionization for the range of densities $x\le 10$, and thus we set $S^N(x,q,T)=1.0$ for $N=\left\{1,2,10\right\}$.
However, at densities $x> 10$, the \citet{summersRAL} data  for these three isoelectronic sequence show suppression factors that are fit well by the usual Eq.~\ref{eqsuppression2}, so we do not  modify $S^N(x,q,T)$ for these cases.

Lastly, we consider the intermediate isoelectronic sequences for which excitation
arises from neither a fine-structure splitting of the ground state nor
a change in principal quantum number of the core.
These include the
Li-like $2s\rightarrow 2p$,
Be-like $2s^2\rightarrow 2s2p$,
N-like $2s^22p^3(^4S)\rightarrow 2s2p^4(^4P)$,
Na-like $3s\rightarrow 3p$,
Mg-like $3s^2\rightarrow 3s3p$,
and P-like $3s^23p^3(^4S)\rightarrow 3s3p^4(^4P)$ cases up through the third row sequences.
Any large low-temperature DR contribution arising from
near threshold resonances is to low-lying states, for which
suppression is negligible, i.e. the high-temperature suppression factor must be switched-off ($S^N \rightarrow 1$) at low-$T$.

\begin{figure}[!hbtp]
\centering
\includegraphics[width=\textwidth]{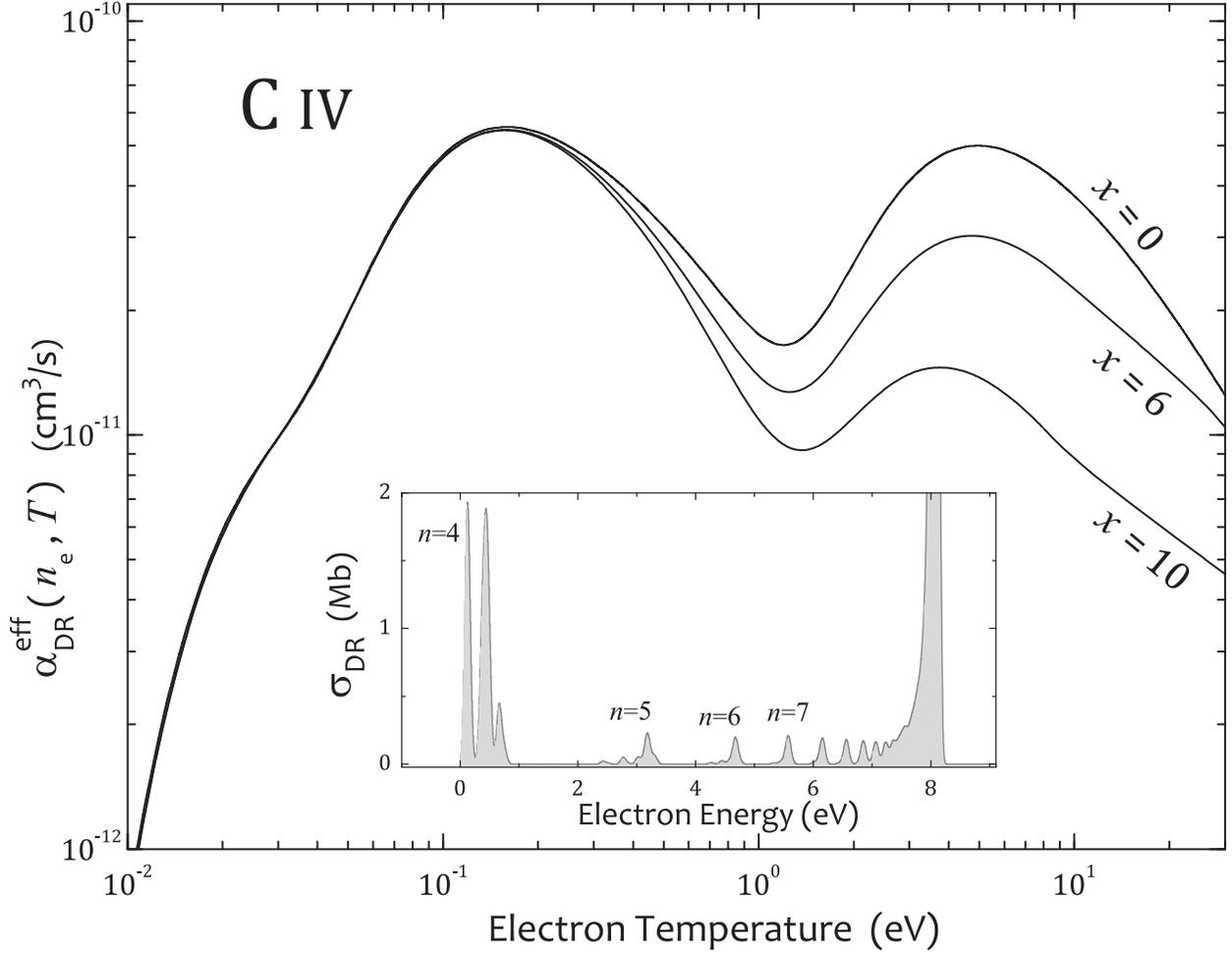}
\caption{\label{fig4}
DR of \ion{C}{4}.
The inset shows the (zero-density) DR cross section convoluted with a 0.1 eV FWHM Gaussian.
The spectrum is dominated by two features: the $n=4$ DR resonance manifold below 1.0 eV and the $n\rightarrow\infty$ Rydberg resonances
accumulating at the $2s\rightarrow 2p$ series limit $\epsilon_{3}(q_0)\approx 8$ eV.
The main figure shows the effective DR rate coefficient for several densities.
Our modified suppression formulation for $x>0$, using Eqs.~\ref{eqsuppression2} and \ref{eqsmod},
ensures that the high-$T$ peak, corresponding to the $n\rightarrow\infty$ Rydberg series of resonances, is
suppressed  but the low-$T$ peak, corresponding to the $n=4$ resonances, is not suppressed.
}
\end{figure}

To illustrate the general demarcation between low-$T$ and high-$T$ DR, we first consider DR of \ion{C}{4}, an overview of which is
depicted in Fig.~\ref{fig4}.  The DR cross section, shown in the inset, is dominated by two features.
 The first is the $n\rightarrow \infty$ accumulation of resonances at the $\epsilon=8$ eV series limit --- those which can be treated
 in the usual high-$T$ fashion \citep{burgess,burgsum} and are therefore susceptible to
 suppression according to our formulation above.
However, there is a second strong contribution from the lowest accessible resonances just above the threshold electron energy, which,
according to the Rydberg consideration $3^2/n^2\approx \epsilon_3=0.6$ Ryd, occur here for $n=4$.
More generally, these low-lying states are typical of the low-lying DR spectrum \citep{nussbaumer}\footnote{
We note that the \ion{C}{4} $n=4$ resonance manifold has been the subject of further
near-threshold density-dependent effects \citep{c4}.}.
The $1s^22p4l$  resonances decay predominantly to the $1s^22s2p$, $1s^22p^2$ and $1s^22s4l$ states. These states lie well below the
ionization limit and so are not susceptible to further reionization. Since there should be no density suppression then,
we seek a modified suppression factor which tends toward unity (i.e., no suppression) at lower temperatures.

In order to make a smooth transition from the high-$T$ suppression factor $S\left(x;q,T\right)$ given in Eq.~\ref{eqsuppression2},
which is appropriate for the high-$T$ peak region $kT\approx kT_{max}=2\epsilon_N/3$, to the low-$T$ region, where $S^N \rightarrow 1$,
we use the modified factor
\begin{eqnarray}
S^N(x;q,T) & = & 1 - \left[ 1 - S\left(x;q,T\right) \right]\, \exp\left(-\frac{\epsilon_N(q)}{10kT}\right) \ ,
\label{eqsmod}
\end{eqnarray}
where $\epsilon_N(q)=8$ eV for the particular case of \ion{C}{4} ($N=3$ and $q=3$).
As seen in Fig.~\ref{fig4}, the density-dependent effective DR rate coefficient,
$\alpha_{DR}^{eff}(n_e,T)$, indeed satisfies the requirement that the high-$T$ peak
is suppressed according to the formulation of \cite{badnell} whereas suppression
is totally turned off for the lower-$T$ peak.

We have investigated the application of Eq.~\ref{eqsmod} for all ions that exhibit these same low-$T$ resonances features, namely,
all isoelectronic sequences $N=\left\{3,4,7,11,12,15\right\}$, and we have found that the correct transitioning
from suppression at the high-$T$-peak to no suppression at low-$T$ is indeed satisfied,
provided, of course, that the appropriate dipole-allowed excitation energy $\epsilon_N(q)$ is employed.
For efficient representation, the excitation energies along each isoelectronic sequence are parameterized by the expression
\begin{eqnarray}
\epsilon_N(q) & = & \sum_{j=0}^5 p_{N,j} \left(\frac{q}{10}\right)^j\ .
\label{eqepsilon}
\end{eqnarray}
These parameters, which are determined by fitting the above expression to the available NIST excitation energies \citep{nist}, are
listed in Table~\ref{table1}.
\begin{deluxetable}{lcrrrrrr}
\tabletypesize{\scriptsize}
\tablecaption{\label{table1}
Fitting coefficients for the excitation energies $\epsilon_N(q)=\sum_{j=0}^{5} p_{N,j} \left(\frac{q}{10}\right)^j$, in eV. Numbers in square brackets denote powers of 10.}
\tablewidth{0pt}
\tablehead{
\colhead{Sequence} & \colhead{$N$} & \colhead{$p_{N,0}$} & \colhead{$p_{N,1}$} & \colhead{$p_{N,2}$} & \colhead{$p_{N,3}$} & \colhead{$p_{N,4}$} & \colhead{$p_{N,5}$}
}
\startdata
Li-like                & 3           & 1.963[+0] &  2.030[+1] & -9.710[-1] &  8.545[-1] & 1.355[-1] &  2.401[-2] \\
Be-like                & 4           & 5.789[+0] &  3.408[+1] &  1.517[+0] & -1.212[+0] & 7.756[-1] & -4.100[-3] \\
~N-like                & 7           & 1.137[+1] &  3.622[+1] &  7.084[+0] & -5.168[+0] & 2.451[+0] & -1.696[-1] \\
Na-like                & 11          & 2.248[+0] &  2.228[+1] & -1.123[+0] &  9.027[-1] & -3.860[-2] &  1.468[-2] \\
Mg-like                & 12          & 2.745[+0] &  1.919[+1] & -5.432[-1] &  7.868[-1] & -4.249[-2] &  1.357[-2] \\
~P-like                & 15          & 1.428[+0] &  3.908[+0] &  7.312[-1] & -1.914[+0] & 1.051[+0] & -8.992[-2] \\
\hline
H-, He-, Ne-like           & 1,2,10        & $\infty^{\dag}$ & 0.0        &  0.0       &  0.0       & 0.0       & 0.0        \\
\hline
B-, C-, O-, F-like     & 5,6,8,9     & 0.0 & 0.0        &  0.0       &  0.0       & 0.0       & 0.0        \\
Al-, Si-, S-, Cl-like  & 13,14,16,17 & 0.0 & 0.0        &  0.0       &  0.0       & 0.0       & 0.0        \\
\hline
                       & $\ge 18$    & 0.0 & 0.0        &  0.0       &  0.0       & 0.0       & 0.0        \\
\hline
& &
\multicolumn{6}{l}{${}^{\dag}$ {\footnotesize Reset to 0.0 for $x>10$. }} \\
\hline
\enddata
\end{deluxetable}

We note that all isoelectronic sequences and ionization stages are  now included in this prescription --- our final comprehensive model
for treating DR suppression, albeit in a simplified fashion.
For those ions with fine-structure splitting in the ground state, we have $\epsilon_N(q)\approx 0$, so that $S^N(x;q,T)=S(x;q,T)$.
(We apply this generally also for Ar-like sequences and above ($N\ge 18$), based-on the density of states --- see, for example, \cite{badnell3pq}.)
For the closed-shell cases,
on the other hand,  we have $\epsilon_N(q)\rightarrow \infty$. Thus, $S^N(x;q,T)=1$ for hydrogenic and closed-shell systems, i.e.,
there is no suppression (for $x\le 10$).  Lastly, for the intermediate cases,
the suppression factor is gradually increased toward unity at lower temperatures and begins to admit
low-$n$ DR resonances.

\section{Results}

The suppression factors derived here have been applied to the state-of-the-art total DR
rate coefficients taken from the most recent DR database.\footnote{\tt http://amdpp.phys.strath.ac.uk/tamoc/DATA/}
These modified data have been incorporated into version C13 of the plasma simulation
code Cloudy, most recently described by \cite{cloudy13}.
Cloudy can do simulations of both photoionized and collisionally ionized cases,
and we show the effects of collisional suppression on both.

\begin{figure}[hbtp]
\centering
\includegraphics[width=\textwidth]{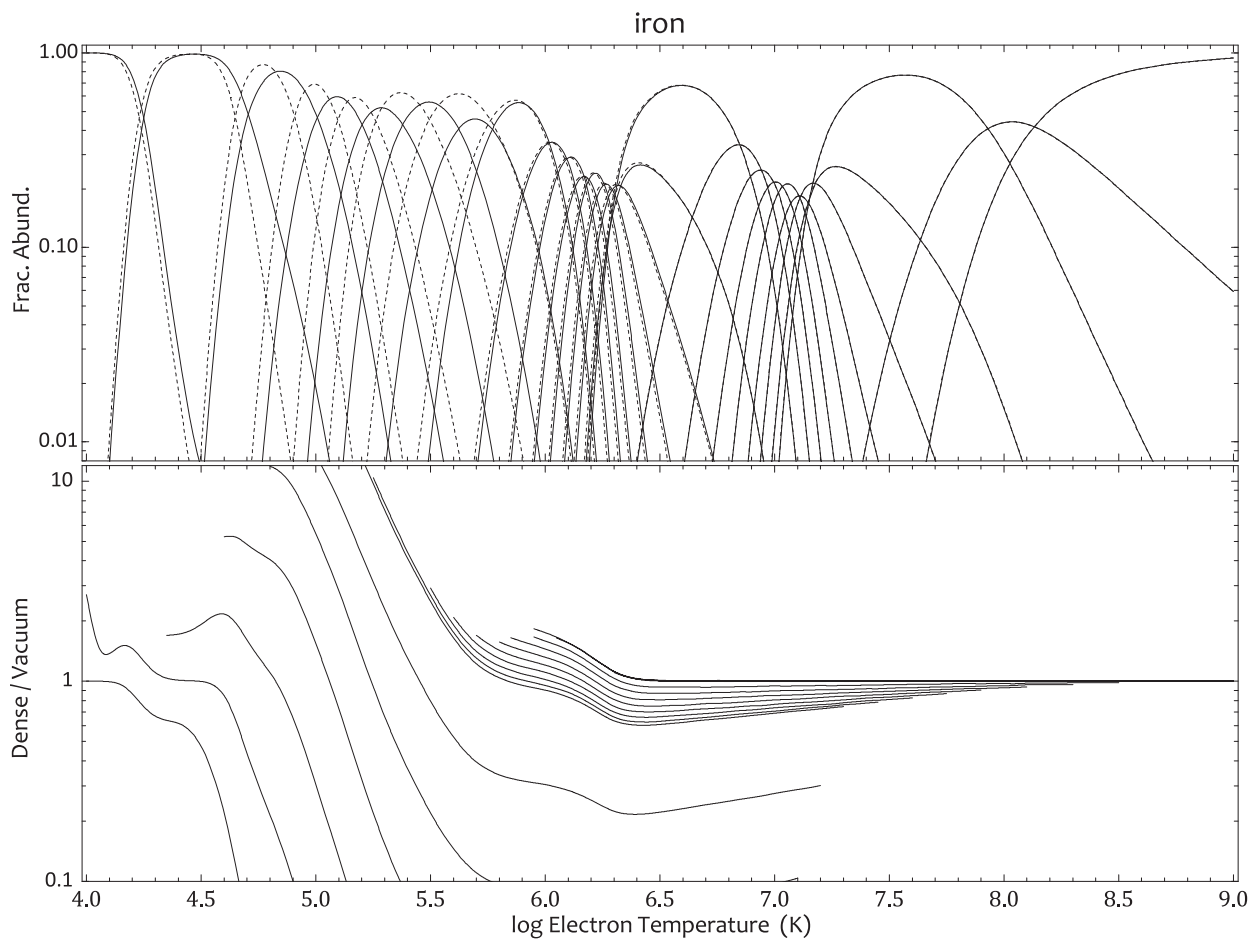}
\caption{\label{figcollision}
Upper panel:
collisional ionization fractional abundance vs. electron temperature for all ionization stages of Fe. The solid curves
 correspond to a density of 1 cm$^{-3}$ and the dashed curves
 correspond to a density of $10^{10}~\rm{cm}^{-3}$.
From left to right, the curves range from \ion{Fe}{1} to \ion{Fe}{27}.
Lower panel: ratio of the calculated fractional abundances for the two densities.
}
\end{figure}

\begin{figure}[hbtp]
\centering
\includegraphics[width=\textwidth]{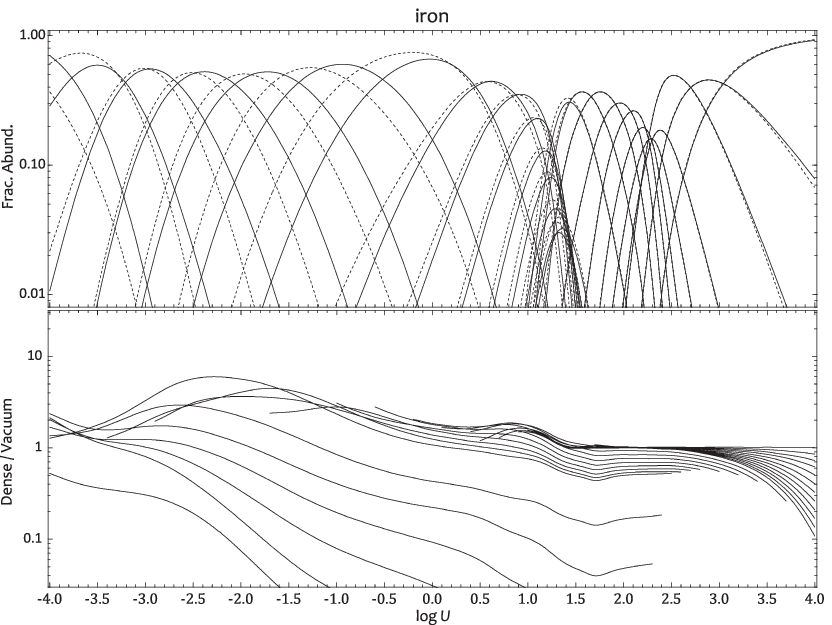}
\caption{\label{figphoto}
Upper panel: photoionization fractional abundance vs. the ionization parameter $U$ for all ionization stages
of Fe.
The solid curves
 correspond to a density of 1 cm$^{-3}$ and the dashed curves
 correspond to a density of $10^{10}~\rm{cm}^{-3}$.
From left to right, the curves range from \ion{Fe}{1} to \ion{Fe}{27}.
Lower panel: ratio of the calculated fractional abundances for the two densities.
}
\end{figure}

Figure \ref{figcollision} shows the ionization distribution of iron for the collisional ionization case.
Figure \ref{figphoto} shows a similar calculation for photoionization equilibrium.
Both show two hydrogen densities, 1 cm$^{-3}$, where collisional suppression of DR should be
negligible, and 10$^{10}$ cm$^{-3}$, where collisional suppression should greatly affect
the rates for lower charges and temperatures.
The upper panel shows the ionization fractions themselves, for these two densities,
while the lower panel shows the ratio of the high to low density abundances.

%\clearpage

Cloudy's assumptions in computing collisional ionization equilibrium, as
shown in Figure \ref{figcollision}, have been described by \citet{lykins}.
It is determined by the balance between collisional ionization from the ground
state and recombination by radiative, dielectronic, and three body recombination to all levels
of the recombined species.

The photoionization case shown in Figure \ref{figphoto} depicts the Active Galactic Nucleus
spectral energy distribution (SED), described by \citet{mathews}, as a function of the
ionization parameter
\begin{eqnarray}
U & \equiv & \frac{\Phi_H}{n_H\;c}\ ,
\end{eqnarray}
where $\Phi_H$ is the hydrogen-ionizing photon flux, $n_H$ is the density of hydrogen, and $c$ is the speed of light.
There is only an indirect relationship between the gas kinetic temperature and the ionization of the gas in this case.
Here, the level of ionization is determined by a balance between  photoionization by the energetic continuum
and the total recombination rate.

The lower panels of Figs.~\ref{figcollision} and \ref{figphoto} show that the amount that the
ionization increases due to DR suppression can be large --- the ratio can easily exceed 1 dex.
Clearly, these results demonstrate that density effects on the ionization balance need to be
considered more precisely.

\section{Conclusion}

We have investigated the effects of finite densities on the effective DR rate coefficients
by developing a suppression factor model, which was motivated by the early work of \citet{badnell} for
\ion{C}{4} and extended to all other ions using physically-motivated scaling considerations,
and more precise fitting of collisional-radiative data \citep{summersRAL}.
Accurate zero-density DR rate coefficients were then multiplied by this suppression factor and introduced
into Cloudy to study the finite-density effects on computed ionization balances of both collisionally ionized and photoionized plasmas.
It is found that the difference in ionization balance between the near-zero and finite-density cases is substantial,
and thus there is sufficient justification for further studies of collisional suppression from generalized collisional-radiative calculations.
This is expected to impact the predictions of the ionization balance in denser cosmic gases such as those
found in nova and supernova shells, accretion disks, and the broad emission line regions in active galactic nuclei.

The present results are intended to be preliminary, and to demonstrate the
importance of density effects on dielectronic recombination in astrophysical plasmas.
Given the approximations adopted, we suggest that their incorporation into models
(e.g., via Cloudy) be used with a little caution. For example, one might run models with and without
the effects of suppression at finite density, especially in modeling higher density plasmas
(e.g., the broad emission line region in quasars).
Nevertheless, it is nearly half a century since \cite{burgsum}
demonstrated significant density effects on DR, and it is time that some representation exists within astrophysical
modeling codes to assess its impact on the much more rigorous demands made by modern day modeling,
especially given its routine incorporation by magnetic fusion plasma modeling codes.
In the longer term, we intend to present results based on detailed
collisional-radiative calculations using state-of-the-art state-specific
DR rate coefficients.

\section{Acknowledgments}

DN, TWG, and KTK acknowledge support by NASA (NNX11AF32G).
GJF acknowledges support by NSF (1108928; and 1109061), NASA (10-ATP10-0053, 10-ADAP10-0073, and NNX12AH73G), and
STScI (HST-AR-12125.01, GO-12560, and HST-GO-12309).
UK undergraduates Mitchell Martin and Terry Yun assisted in coding the DR routines used here.
NRB acknowledges support by STFC (ST/J000892/1).

%===============================%
%===% SECTION: BIBLIOGRAPHY  %===%
%===============================%
%%\bibliographystyle{aa}
%%\bibliography{sdr}

\end{document}